# A scanning tunneling microscopy based potentiometry technique and its application to the local sensing of the spin Hall effect

Ting Xie[1, a)], Michael Dreyer[2], David Bowen[3], Dan Hinkel[3], R. E. Butera[3], Charles Krafft[3], and Isaak Mayergoyz[1]

[1]Department of Electrical and Computer Engineering, University of Maryland, College Park, Maryland 20742, USA
[2]Department of Physics, University of Maryland, College Park, Maryland 20742, USA
[3]Laboratory for Physical Sciences, College Park, Maryland, 20740, USA

Abstract

A scanning tunneling microscopy based potentiometry technique for the measurements of the local surface electric potential is presented and illustrated by experiments performed on current-carrying thin tungsten films. The obtained results demonstrate a sub-millivolt resolution in the measured surface potential. The application of this potentiometry technique to the local sensing of the spin Hall effect is outlined and some experimental results are reported.

Recent technological progress has resulted in the reduction of the dimensions of electronic and magnetic devices to the nanoscale. One of the many challenges associated with this nanoscale reduction is to develop reliable techniques for the measurement of local electrical properties of materials. To handle this problem, various techniques have been proposed to measure the local surface potential at the nanoscale and used in studies of graphene and other materials.[1-4]

In this letter, we describe a non-contact surface potential measurement (potentiometry) technique by employing the scanning tunneling spectroscopy (STS) function of a scanning tunneling microscope (STM). Since the STS is usually an embedded function in common STM systems, this potentiometry technique hence requires no further modification of an STM system. This makes the technique readily accessible for virtually all STMs. It is demonstrated below that this potentiometry technique has a sub-millivolt resolution of surface potential measurements and nanometer resolution in positioning. As a result, the resistivity of conducting films (and its uniformity) can be fully characterized at the nanoscale. Furthermore, this potentiometry technique is applied to the STM based study of the spin Hall effect (SHE) in tungsten films and some experimental results of this study are reported.

To illustrate the potentiometry technique by using an STM, 5 nm-thick tungsten films on sapphire substrates have been fabricated by using DC magnetron sputtering and analyzed

a) Email: tingxie@terpmail.umd.edu

*in-situ* under ultra-high vacuum conditions. A schematic diagram for the potentiometry analysis is presented in Fig. 1. This figure shows that one of the terminals of the sample is connected to the voltage source while the other terminal is connected to a current source to control a tunneling gap voltage ($V_g$) and supply a desired bias current ($I_{bias}$) through the tungsten film, respectively.

The flow of the bias current results in a voltage drop along the tungsten film. To detect the potential at any ($x, y$)-location on the sample, we have implemented the following measurement procedure based on STS. The measurement starts by parking the STM tip at the desired ($x, y$)-location. Then, the feedback loop of the STM system is turned off to freeze the sample and tip at their desired positions, i.e. to eliminate any possible mechanical movements. Next, the tunneling current $I_t$ is measured at different values of the source voltage $V_s$. After the measurement of $I_t$-$V_s$ curve is performed, the feedback loop is switched on and the tip can be moved to other locations for further measurements of the potential. In our experiments, the terminal of the sample connected to the voltage source is used as the reference point for the surface potential. The bias-current-induced surface potential is determined from the obtained $I_t$-$V_s$ curves by applying Kirchhoff's voltage law (KVL) to the loop indicated by the blue dash line in Fig. 1. Since the STM tip is virtually grounded by the pre-amplifier, the KVL equation for the above loop can be written as follows

$$V_s + V(x, y) = V_g + I_t R_{tip}, \qquad (1)$$

where $V_s$ is the source voltage, $V(x, y)$ is the voltage between the tunneling location and the reference point, $V_g$ is the tunneling gap voltage, $I_t$ is the tunneling current, and $R_{tip}$ is the tip resistance. It is apparent that a tunneling current of zero corresponds to zero tunneling gap voltage. Therefore, the right-hand side of equation (1) is equal to zero when $I_t = 0$. In this case, the surface potential $V(x, y) = -V_s$, where the value of $V_s$ at which $I_t = 0$ can be found by locating the intersection of the $I_t$-$V_s$ curve with the $I_t = 0$ line.

Typical $I_t$-$V_s$ curves obtained at $I_{bias} = 0$ mA, 0.2 mA and -0.1 mA are shown in Fig. 2a. By identifying the intersections of these curves with the $I_t = 0$ line, the surface potentials due to the above bias currents are found to be 0 mV, 52 mV, and – 26 mV, respectively. The surface potential at any fixed location is a linear function of $I_{bias}$. Hence, by applying linear regression to the data shown in the inset, we found that the resistance is about 260.7 Ω between the tunneling location and the reference point. Surface potentials measured at different ($x, y$)-locations on the tungsten thin film at a constant bias current reveal the resistivity of the tungsten film. The results of these measurements are illustrated in Fig. 2b. The potential gradient in the x-direction is found to be 1.20 mV/μm at $I_{bias} = 10$ mA. In fact, the good linear fit of the surface potentials measured at various locations indicates a good uniformity in the thickness of the deposited film. Since the x-direction is parallel to the direction of the current flow, the resistivity of the film can then be estimated to be 240 μΩ·cm for the tungsten film with a cross-section of 5 nm × 4 mm. This high resistivity is a fingerprint of a β-phase tungsten film and it is consistent with results reported in the literature.[5,6] As expected, the potentiometry measurements along the y-axis at $(x, y)$ = (0 µm, 0 µm), (0 µm, -0.6 µm), and (0 µm, 0.6 µm) show no potential differences because the y-direction is perpendicular to the current flow direction.

To illustrate the lateral resolution of the described potentiometry technique, measurements have been performed at the location of a gold nanoparticle deposited on the tungsten film through a shadow mask, as shown in Fig. 3. The STM image of the nanoparticle (Fig. 3a) shows a layer-by-layer structure of the deposited gold. The surface potential distribution in Fig. 3b was obtained for 10 mA bias current flow in the x-direction. This measurement demonstrates the nanometer scale lateral resolution of the described potentiometry technique. Indeed, the image shows a clear distortion in the distribution of the surface potential on the tungsten film due to the gold particle. This distortion is evident from the abrupt drop in surface potential in the vicinity of the "current-encountering" part of the particle boundary (right-hand side). Meanwhile, the potential difference between the gold area and the tungsten film on the left-hand side of the gold particle is so small that a clear boundary cannot be identified. The features of the potential map of the gold nanoparticle are consistent with its high conductivity.

Next, we demonstrate that the described potentiometry technique is instrumental for an STM based study of the spin Hall effect (SHE) in conducting films, which is currently of great interest in spintronics research. The SHE manifests itself in the accumulation of the spin-polarized electrons on the surface of a current-carrying sample.[5-8] In the STM study of the SHE, it is very important to maintain a constant tunneling gap voltage between the tip and the conducting film in the presence of current flow through the film, which can be achieved by using the potentiometry technique. Indeed, by neglecting the very small term $I_{tip}R_{tip}$ (~ nV) in equation (1), we find that a desired tunneling voltage $V_g$ can be accomplished by setting the source voltage $V_s = V_g - V(x, y)$. With this technique, we have performed an STM study of the SHE by using tungsten films and tungsten tips. In the study, the tunneling current has been measured in the presence of a current pulse (~ 1 ms) at a constant tunneling voltage of 0.5 V with the STM feedback turned off to freeze the position of the tip and sample. In our experiments, we observed the following two distinct phenomena. First, there is a substantial and gradual in time increase in the tunneling current, as shown in Fig. 4a. This can be attributed to a change in the tip-sample distance (tunneling gap) in the presence of the bias current through the tungsten film. These gradual tunneling gap changes have been confirmed by using atomic force microscopy (AFM) (see inset of Fig. 4a, where the results of AFM measurements are presented). There, the shift of the cantilever resonant frequency (*df*) is caused by the decrease in the tunneling gap during the $I_{bias}$ pulse. These tunneling gap changes are most likely caused by local thermal expansion. Second, we have also observed an asymmetry in the tunneling currents caused by switching the polarity of the tunneling gap voltage in the presence of bias current pulses, as shown in Fig. 4b. This figure shows the difference in the tunneling currents measured independently at -0.5 V and 0.5 V with identical bias current flows through the film. The difference with no bias current (0 mA) is almost zero while the differences increase in time in the presence of 5 mA and 7 mA bias currents. This temporal increase in the difference is due to the increase in the tunneling current caused by the bias current as discussed for Fig 4a. Therefore, the temporal increase can be eliminated by normalizing the difference in the tunneling currents measured at -0.5 V and 0.5 V to their mean values. The normalized results are shown in the inset of Fig. 4b, which clearly indicate a bias-current-induced asymmetry in the tunneling currents measured at different polarities of the tunneling voltages. This asymmetry can be explained by the asymmetry in the tunneling process caused by the SHE. Indeed, on the one hand, when the polarity of the tunneling gap voltage is such that electrons tunnel from the tungsten film, we deal with the tunneling of electrons which are spin-polarized due to the SHE. On the other hand, when the polarity of the tunneling voltage is such that electrons tunnel from the tip to the film, we deal with the tunneling of non-spin-polarized electrons. The fact that this asymmetry is enhanced by the increase in the bias current through the film (see the inset of Fig. 4b) also supports that this mentioned asymmetry is related to the SHE. The described experimental results suggest that the local STM sensing of the SHE in conducting films can be achieved with the potentiometry technique.

In summary, we have described and experimentally illustrated a potentiometry technique for the measurement of the surface potential in a conducting film with a common STM system. The unique feature of this technique is the use of the built-in STS function to detect the bias-current-induced surface potential. In this way, an STM system can be easily configured to characterize the resistivity of a sample and its thickness uniformity at the nanoscale. It is also demonstrated that this potentiometry technique is instrumental for the local STM study of the SHE. It is also apparent that the presented potentiometry technique opens the opportunity for the study of transport phenomena at the nanoscale in conducting and semiconductor films.

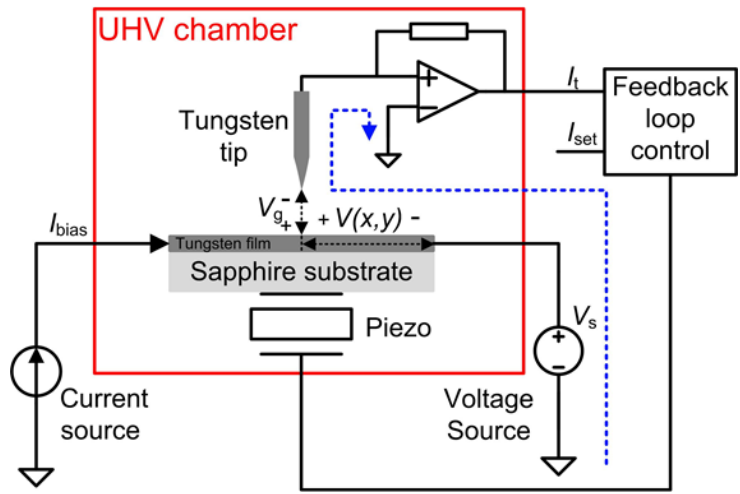

Figure 1. Schematic diagram of the potentiometry analysis with an STM. The blue dash line indicates the loop to which Kirchhoff's voltage law was applied.

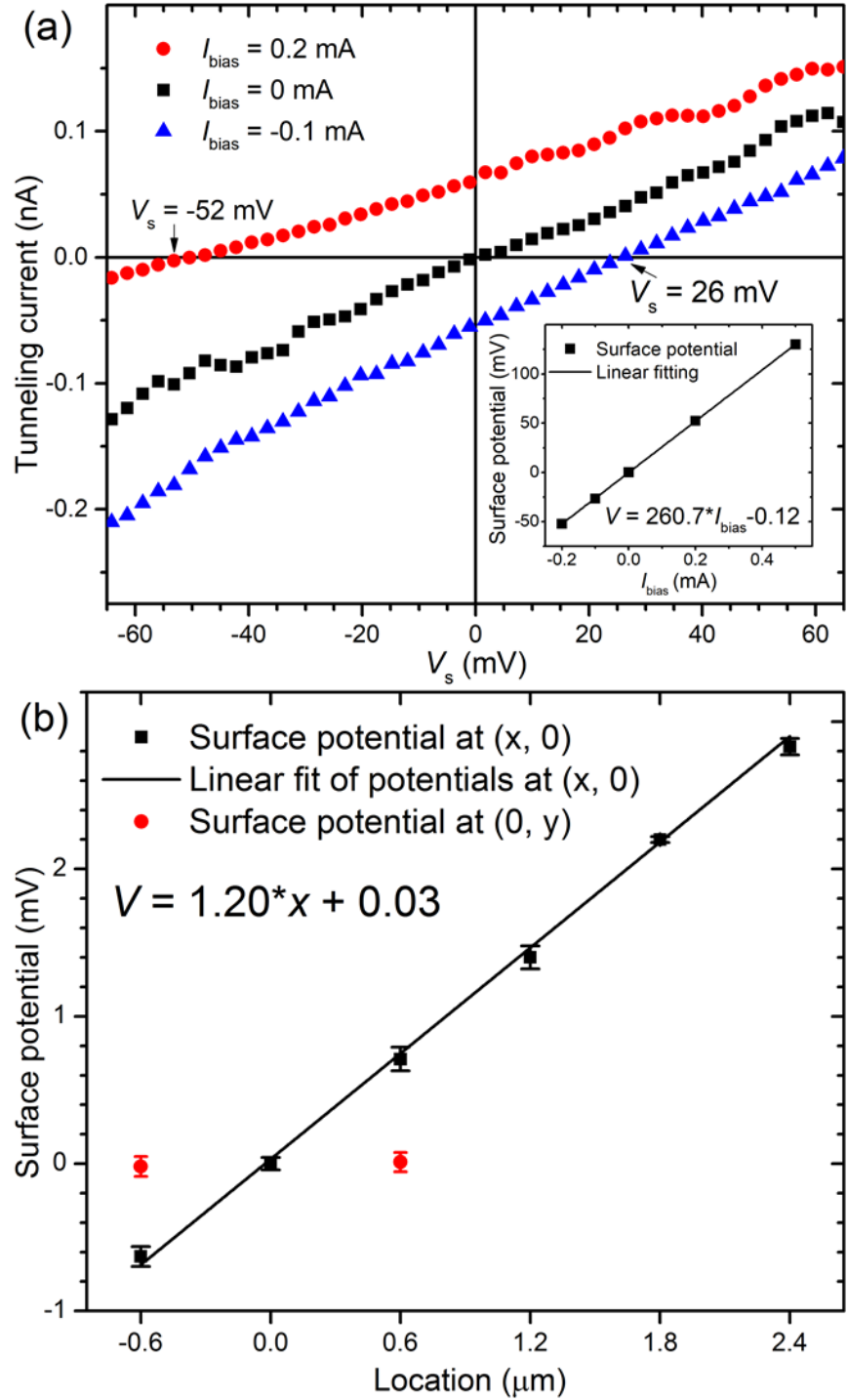

Figure 2. (a) $I_t$-$V_s$ curves obtained at one location on the tungsten film with $I_{bias}$ = 0 mA, 0.2 mA and -0.1 mA. Inset: The surface potential obtained with different bias currents at the same location. The linear fit of the data shows a slope of 260.7 V/A. (b) Surface potentials measured at various locations with a bias current of 10 mA. The slope of the linear fitted line of surface potentials measured in the x-direction is 1.20 mV/μm.

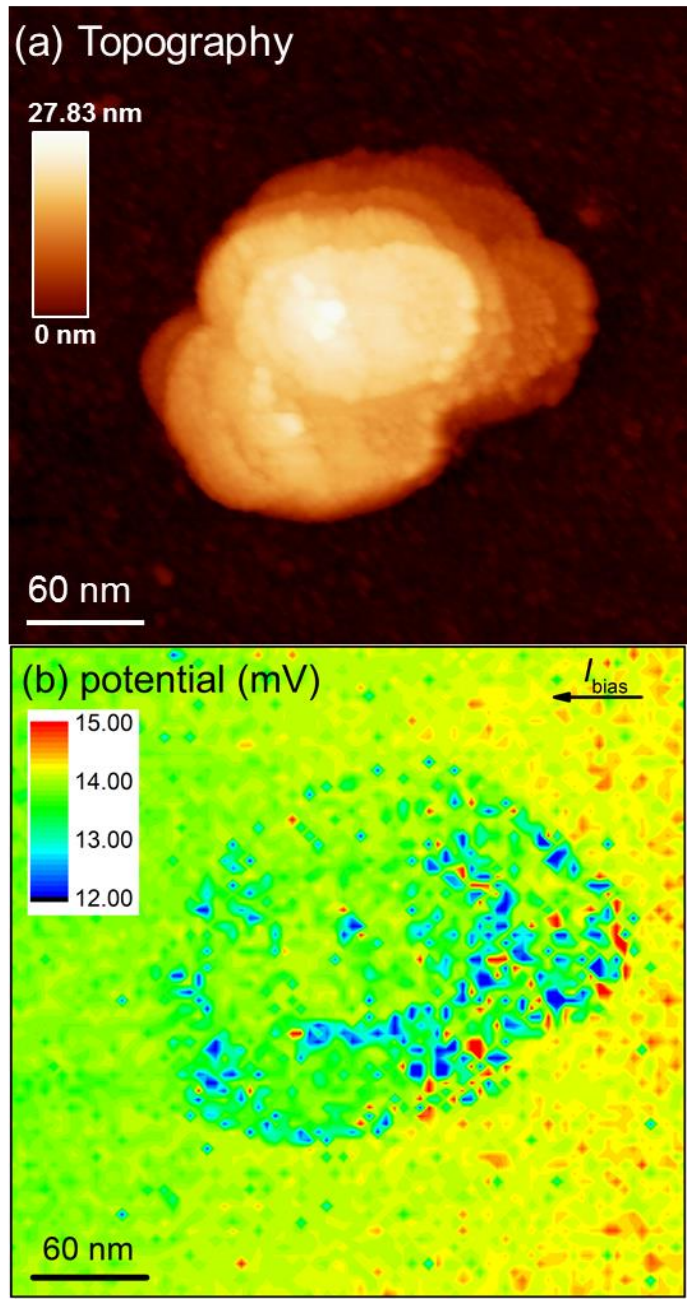

Figure 3. (a) STM topography and (b) surface potential map of a gold nanoparticle on the tungsten film. The bias current is 10 mA with the direction indicated in (b).

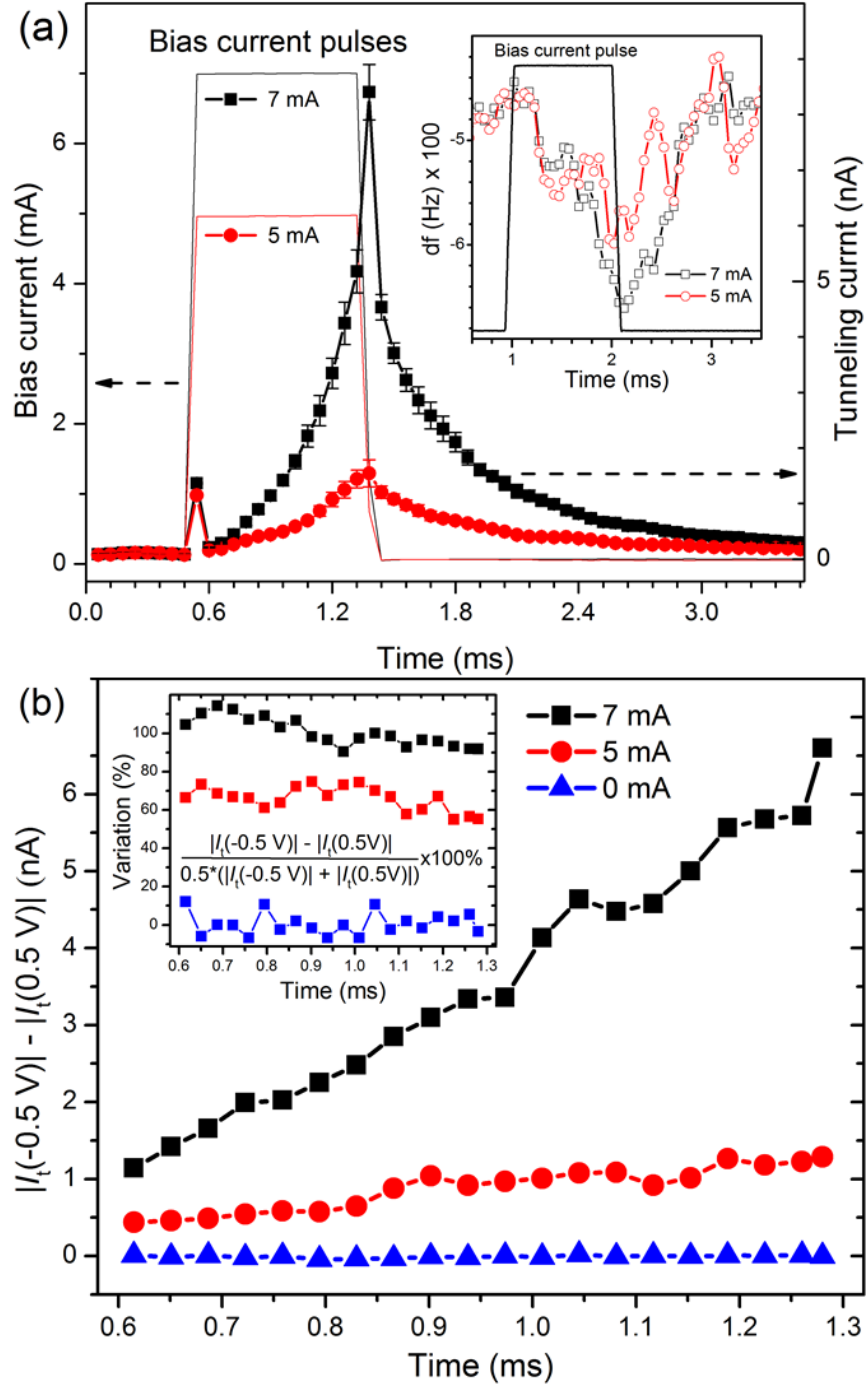

Figure 4. (a) Tunneling currents in the presence of bias current pulses with different amplitudes. The tunneling gap voltage was maintained at 0.5 V for all the measurements. Inset: atomic force microscopy study on the tungsten film in the presence of bias current pulses. (b) Difference in tunneling currents measured at tunneling voltages of -0.5 V and 0.5 V in the presence of identical current pulses. Inset: the normalized difference to the mean of the tunneling currents measured at 0.5V and -0.5 V.